\documentclass[a4paper,12pt]{article} 

\usepackage[english]{babel}
\usepackage{authblk}
\usepackage{dsfont}
\usepackage{amsfonts}
\usepackage{mathrsfs}
\usepackage{amssymb}        
\usepackage{amsmath}
\usepackage{graphicx}
\usepackage{float}
\usepackage[a4paper]{geometry} 
\usepackage{verbatim}
\usepackage{pstricks}
\usepackage{amsthm}
\usepackage{mathrsfs}

\newcommand{\Z}{{\mathbb Z}}
\newcommand{\R}{{\mathbb R}}
\newcommand{\dd}{{\mathrm d}}

\newcommand{\ed}{{\mathrm e}}

\usepackage[margin=1cm,font=small]{caption}

\begin{document}



\author[*]{Fran\c cois Huveneers}
\affil[*]{\small{Universit{\'e} Paris-Dauphine, PSL Research University, CNRS, CEREMADE, 75016 Paris, France}}

\date{\today}

\title{Classical and quantum systems: transport due to rare events}                              

\maketitle

\bigskip

\begin{abstract}
\noindent
We review possible mechanisms for energy transfer based on  `rare' or `non-perturbative' effects, in physical systems that present a many-body localized phenomenology.
The main focus is on classical systems, with or without quenched disorder. 
For non-quantum systems, the breakdown of localization is usually not regarded as an issue, and we thus aim at identifying the fastest channels for transport. 
Next, we contemplate the possibility of applying the same mechanisms in quantum systems, 
including disorder free systems (e.g.\@ Bose-Hubbard chain), disordered many-body localized systems with mobility edges at energies below the edge, and strongly disordered lattice systems in $d>1$. 
For quantum mechanical systems, the relevance of these considerations for transport is currently a matter of debate. 
\end{abstract}

\newpage
\section{Introduction}

Based on the basic principles of thermodynamics, one expects usual pieces of material to eventually reach a state that has a maximal entropy compatible with the macroscopic conservation laws of the system 
(like energy, momentum, number of particles\dots).
This process is referred to as thermalization, since the finial state is characterized by a uniform temperature. 
The microscopic description of this phenomenon relies usually on the ergodic hypothesis. 
For classical systems, this refers to the fact that, for a typical initial state with respect to the Gibbs state (parametrized by the conjugated variables to each macroscopically conserved quantity),
the (infinite)-time average of a local observable approaches the value of this observable in the Gibbs state, as we take the thermodynamic limit.
For quantum systems, the content of the ergodic hypothesis is very much the same: 
In the thermodynamic limit, the value of a local observable in a given eigenstate coincides with its value in the thermal state at the temperature corresponding to the energy of the state (assuming that there is only one conserved quantity).
This formulation is known as the eigenstate thermalization hypothesis (ETH) \cite{Deutsch}\cite{Srednicki}. For quantum systems, we will use the words `ergodic' and `ETH-obeying' interchangeably. 

Early observations of mechanical systems with a few degrees of freedom raised doubts on the validity of the ergodic hypothesis.   
E.g.\@ a surprising lack of thermalization was observed for a classical chain made of 64 atoms in the celebrated Fermi-Pasta-Ulam (FPU) experiment \cite{FermiPastaUlam}. 
Later, it was shown at the mathematical level of rigor, through the Kolmogorov-Arnold-Moser (KAM) theorem \cite{Kolmogorov,Arnold_1963,Moser}\cite{Poschel_2001}, 
that a positive fraction of the phase space of generic non-integrable classical systems with a finite number of variables, can be occupied by quasi-periodic trajectories. 
This result was naturally invoked to explain the upshot of the FPU experiment, despite important discrepancies between the strength of the interactions in KAM and FPU.

However, these facts do not directly concern the thermodynamic limit, and could simply be considered as important finite size effects, as is nowadays most commonly believed to be the case for the FPU chain.
Anderson localization instead implies a complete lack of equipartition of energy in the thermodynamic limit. 
This phenomenon was initially argued to hold for non-interacting electrons by Anderson \cite{Anderson}, 
and the stability of localization in the presence of interactions was studied in subsequent works \cite{FleishmanAnderson,GornyiMirlinPolyakov,BAA}.
In \cite{BAA}, the authors conclude at a persistence of localization for weak enough interactions and low enough temperatures. 
The existence of a many-body localized (MBL) phase shows thus that the ergodic hypothesis is clearly invalidated for a generic class of quantum systems.

A detailed description of MBL, in terms of local integral of motion, is available for the following systems: strong disordered, one-dimensional, fermionic or spin chains, where the full spectrum is localized 
\cite{OganesyanHuse,HuseNandkishoreOganesyan,SerbynPapicAbanin_2013,Imbrie_JSP,Imbrie_PRL}. 
For systems not in this class, the possibility of MBL is either debated, either known to not exist.  
This is most notably the case for classical systems \cite{Basko}. 
Indeed, just a few coupled oscillators contain chaotic islands for any non-trivial values of the parameters in the Hamiltonian \cite{Chirikov}. 
They behave as perfect baths and allow the redistribution of energy in the system (more precisely, in the fraction of phase space not occupied by KAM tori), through a process known as Arnold diffusion \cite{Arnold_1964}. 
The probability, with respect to the Gibbs state at positive temperature, of being on a quasi-periodic orbit is itself believed to vanish in the thermodynamic limit, 
since the probability of having a chaotic configuration somewhere in space tends to one. 

The above discussion focussed on the difference between ergodic systems, and systems that truly break ergodicity, but completely ignored the question of time scales.
Many systems that are not localized may still enjoy some MBL phenomenology, characterized e.g.\@ by transport coefficients that become much smaller than the bare values of the coupling constants appearing in the Hamiltonian.
The study of such cases is important since 
a) the class of systems with an MBL phenomenology may be much broader compared to the class of truly MBL systems, 
b) the difference between approximate and true MBL may be irrelevant for many practical purposes, in the same way as a window glass appears as completely frozen on human time scales. 

The MBL phenomenology can usually be seen as a consequence of some perturbative analysis, that can potentially be destroyed once `rare' or `non-perturbative' phenomena are taken into account.
For example, in the disordered classical spin chain studied in \cite{OganesyanPalHuse}, there is typically some frequency mismatch between near spins, that will drastically slow down the transfer of energy at strong disorder \cite{Huveneers};  
and the transport will ultimately be mediated by some rare chaotic spots that do not show up in any perturbative analysis (the perturbative parameter is here the coupling strength divided by the disorder strength).   
For classical systems with finitely many degrees of freedom, Nekhoroshev estimates  
guarantee that any transport must vanish as a stretch exponential in the perturbative parameter as this parameter goes to zero \cite{Nekhoroshev}\cite{Poschel_1993}. 
These estimates were generalized in \cite{BenettinFroehlichGiorgilli} for systems with infinitely many degrees of freedom and a finite excitation above the ground state 
(the stretch exponential decay is however replaced by the much slower decay of the type $\ed^{- c (\log \epsilon^{-1})^p}$ for some $c>0$ and some $p> 1$, where $\epsilon>0$ is the dimensionless perturbative parameter). 
These estimates are known to be sharp in some cases, and one is naturally led to ask what they become in the thermodynamic limit at positive temperature. 

The aim of this paper is to review what is currently known or conjectured about transport mechanisms due to rare, non-perturbative events, in systems where some MBL phenomenology is present. 
Our main focus is on classical system (Section \ref{sec: classical}), with a particular attention to the work by Basko \cite{Basko} on the thermal conductivity of the disordered non-linear Schr\" odinger (DNLS) equation. 
He identified two possible ways in which the system may thermalize: 
a) Big and abundant fixed chaotic spots influence the rest of the system and 
b) Smaller and less abundant mobile chaotic spots modify the energy landscape as they move accros the system. 
He concluded that the second mechanism is more efficient and governs the transport properties leading, very roughly speaking, to transport coefficients that vanish as $\ed^{- c (\log \epsilon^{-1})^p}$, 
for some $c>0$ and some $p>1$, and for $\epsilon>0$ a dimensionless perturbative parameter of the model (in agreement with the rigorous upper bound in  \cite{BenettinFroehlichGiorgilli} at `zero temperature'). 
We provide an account of this theory that is a bit more detailed than what could be expected in such a review, both because classical systems may be not so familiar to the MBL community, 
and because we want to adapt the arguments of \cite{Basko} to a slightly different, disorder free system, so as to show the universality of these considerations (and actually to simplify slightly the analysis). 
Next, we analyse whether the two above mechanisms could apply to quantum systems as well (Section \ref{sec: quantum}), despite the absence of truly chaotic spots. 
We limit ourselves to a cruder discussion, mainly based on \cite{DeRoeckHuveneers_Scenario,DeRoeckHuveneersMuellerSchiulaz} and \cite{DeRoeckHuveneers_2016}. 
Scenario b) may be realized in disordered free systems at all energies, and in disordered systems with mobility edges \cite{BAA}, for energies below the edge 
(thus, to be more consistent, this scenario implies that a true many-body mobility edge does not exist as such, and must be replaced by a cross-over between a low-energy sector with an MBL phenomenology, and a high energy sector without). 
Scenario a) is actually more delicate to establish for quantum systems, and coud apply to strongly disordered systems in $d>1$ and to systems with subexponentially decaying interactions.   
According to \cite{DeRoeckHuveneers_Scenario,DeRoeckHuveneersMuellerSchiulaz,DeRoeckHuveneers_2016},
all these systems are thus not be truly MBL, i.e.\@ transport does not vanish in the thermodynamic limit, and the dynamics cannot be described by local integrals of motion (not even in the form of \cite{ChandranPalLaumannScardiccio}), 
though some important MBL phenomenology definitely remains.
All this is currently debated, see e.g.\@ \cite{AgarwalAltmanDemlerGopalakrishnanHuseKnap} in the present volume. 

Finally, we want to mention two related topics that this review is not about. 
First, transport in one-dimensional disordered systems can be dominated by rare bottlenecks that can suppress transport very much,  
in the same way as electrical current is slowed down by the strongest resistance in a series circuit.  
This may lead to an important slow-down of transport, becoming possibly sub-diffusive, in the thermal phase, now dubbed the Griffiths phase 
\cite{AgarwalGopalakrishnanKnapMuellerDemler,GopalakrishnanAgarwalDemlerHuseKnap,VoskHuseAltman}. 
This topic is already largely covered in \cite{AgarwalAltmanDemlerGopalakrishnanHuseKnap} and we will not be concerned with it
(though an important part of the work in \cite{Basko} was actually devoted to show that one-dimensional classical systems are typically diffusive).
Second, generic ergodic quantum systems may behave as very bad conductors under the action of some frequency driving: The rate of absorption of energy decays exponentially with the driving frequency
\cite{DalessioPolkovnikov,DalessioRigol,arXivAbaninDeRoeckHuveneersHo_1,arXivAbaninDeRoeckHuveneersHo_2,AbaninDeRoeckHuveneers}. 
A similar phenomenon may be observed in cases where a large energy barrier must be overcome for two sub-systems to interact, as it is the case between singlons and doublons in the Fermi-Hubbard chain
\cite{SensarmaEtAl,StrohmaierEtAl,arXivAbaninDeRoeckHuveneersHo_1,ParameswaranGopalakrishnan}.  
These examples of bad conduction in ergodic systems are probably more fundamental than what we will discuss here, but a detailed treatment would lead us too far from our concerns.

\section{Classical systems}\label{sec: classical}
We consider here two systems that are \emph{a priori} potential candidates to exhibit many-body localization: a disordered and a disorder free classical system of coupled oscillators. 
First we show that indeed, transport is strongly suppressed in some regime of the parameters. This fact results from a perturbative analysis of the conductivity. 
Second, we take non-perturbative phenomena into account, constituted here by chaotic spots, and we show that transport cannot truly vanish in these classical systems. 
This confirms what is probably the most common expectation: 
Localization is not possible in generic classical systems since a finite number of oscillators can be in a configuration where they form a perfect bath 
(presence of chaos or continuous spectrum).

\subsection{Models}
As a prominent example of disordered system, let us consider the disordered non-linear Schr\"odinger (DNLS) system: 
\begin{equation}\label{DNLS}
H (\psi, \overline{\psi}) \; = \; \sum_{x} \big(   \omega_x |\psi_x|^2  + J |\psi_x|^4  \big) + g \sum_{x\sim  y} |\psi_x - \psi_y|^2,
\end{equation}
where $x$ are points on a regular lattice, $g$ is the strength of the hopping, $J \ge 0$ is the strength of the interaction, 
and $\omega_x$ are independent random variables, with $\langle (\omega_x - \langle \omega_x \rangle )^2 \rangle =: W^2$ (and say $\langle \omega_x\rangle = 0$).
This is the model studied by Basko \cite{Basko} at positive temperature in $d=1$ (see below for much more references, though the huge majority of theoretical works deal with a finite excitation above the ground state, 
i.e.\@ a set-up where all oscillators but a few ones are intially at rest). 
The equations of motion are given by the Schr\"odinger equation 
$$i \frac{\dd \psi_x }{ \dd t} \;  = \; \frac{\partial H }{ \partial \overline{\psi}_x} \; = \; \omega_x \psi_x - g (\Delta \psi)_x + J |\psi_x|^2 \psi_x.$$ 
For $J=0$, this equation is linear and the system reduces to an Anderson insulator: all modes are exponentially localized for $g/W$ large enough. 
When $J>0$, the quartic interaction introduces a coupling between the different modes and the question of the persistence of localization arrises naturally.
Besides the energy $H$, the dynamics preserves another extensive quantity: the total norm, also called action  in \cite{Basko}, $N = \sum_x |\psi_x|^2$.
Let $\rho = N/\text{Vol}$. 
The most favorable conditions for localization to be preserved are 
\begin{equation}\label{parameters DNLS}
\frac{g}{W} \; \ll \; 1 \qquad \text{and} \qquad  \tilde\rho \; :=  \; \frac{J \rho}{W} \;\ll \; 1.
\end{equation}
Here we wrote the condition $(J/W) \rho \ll 1$ instead of the more obvious looking $J/W \ll 1$, since the equations of motion are invariant under the scaling $J' = J/\lambda^2$, $\psi_x' = \lambda \psi_x$. 
Finally, we notice that a smallness condition on $J$ is natural if one wishes to consider the interaction as a perturbation. 
However, as we will see in the next example, a very strong interaction may as well favor localization. 

Coupled rotors constitute an example of disorder free system where a tendency to localization can be expected: 
\begin{equation}\label{rotors}
H(p, \theta)  \; = \; \frac{U}{2} \sum_{x} p_x^2 - J \sum_{x \sim y} \cos (\theta_{y} - \theta_x),
\end{equation}
where the variables $p_x$ are canonically conjugated to $\theta_x\in \R\backslash 2 \pi \Z$: 
$$\dot{\theta}_x \; = \; U p_x, \qquad \dot{p}_x \; = \; J\sum_{y:y\sim x} \sin (\theta_y - \theta_x).$$
There are again two conserved quantities: the total energy $H$, and the total momentum, or angular velocity, $O = U \sum_x p_x$. 
However, changing $O$ corresponds to a Galilean boost, and we may assume $O = 0$.  
When $J=0$, the system is trivially localized, and each rotor oscillates at its own frequency $\omega_x = U p_x$. 
Hence, if the system is at equilibrium at temperature $T$, then $\langle p_x^2 \rangle_T = T/U$ and the typical frequency mismatch between near oscillators is $\sqrt{TU}$. 
Therefore, for small interaction strength $J$, a computation at first order in perturbation yields 
$$p_x (t) \; \simeq \;  p_x (0) + J \sum_{y: y\sim x} \int_0^t \dd s \, \sin \big( (\omega_x - \omega_y) s \big),$$
where $\omega_z = \omega_z (t=0)$ for $z=x,y$.  
The integral over time will remain small for all times if 
\begin{equation}\label{localization rotators}
\frac{J}{|\omega_x - \omega_y|} \; \sim \; \frac{J}{\sqrt{UT}} \; \ll \;  1.
\end{equation}
Thus, the quench disorder on $\omega_x$ in the DNLS system is replaced here by a disorder resulting from thermal fluctuations, and it is interesting to notice that higher temperatures favor the localization phenomenon. 
The investigation of an MBL phenomenology in classical disorder free systems seems to have mostly received attention in \cite{Carati,DeRoeckHuveneers_C} (see the next section for quantum systems).

\subsection{Upper-bounds on transport}
All recent numerical studies seem to indicate that classical systems at positive temperature do thermalize \cite{MulanskiAhnertPikovskyShepelyansky,DharSaito,DharLebowitz,OganesyanPalHuse}. 
E.g.\@ in \cite{DharLebowitz}, the authors study an Anderson insulator perturbed by some non-linear term and conclude that `even a small amount of anharmonicity leads to a $\mathcal{J} \sim 1/L$ dependence'
where $\mathcal{J}$ is the current in the non-equilibrium stationary state, and $L$ the linear size of the system (this corresponds thus to normal transport). 
An interesting aspect of \cite{OganesyanPalHuse}, where the diffusivity of a disordered classical spin chain in equilibrium at positive temperature is investigated, 
is the observation that the diffusivity decays drastically with the strength $h$ of the interaction between near spins, `consistent with the possibility that the transport is actually essentially non-perturbative in $h$'. 

The non-perturbative character of the transport can be justified mathematically to a large extent, though not entirely. 
For a final excitation above the ground state, Nekhoroshev type estimates were obtained in  \cite{BenettinFroehlichGiorgilli}, if the total amount of energy above the ground state is small enough (i.e.\@ not just finite), 
as already mentioned in the introduction. 
Similar conclusions were reached in \cite{FishmanKrivolapovSoffer_2008,FishmanKrivolapovSoffer_2009} (see also \cite{FishmanKrivolapovSoffer_2012}), 
where it was also argued that the series expansions must be asymptotic (hence an initially localized wave packet should in general spread over time, for all non-trivial values of the coupling).
See also \cite{WangZhang} for bounds on the diffusion in the DNLS chain. 
The conductivity at equilibrium at positive temperature was studied in \cite{Huveneers} for a large class of disordered systems, and in \cite{DeRoeckHuveneers_C} for disorder free systems, 
the latter being technically more complicated to deal with since resonances have no preferred locations (see also \cite{DeRoeckHuveneers_ProceedingsPSPDE}). 

To see what kind of approximation needs to be made, let us consider the transport of the density $\rho$ in the DNLS system \eqref{DNLS}. 
The conductivity in equilibrium is defined by the Green-Kubo formula 
\begin{equation}\label{Green Kubo}
\kappa (\rho) \; = \; \frac{1}{\rho^2}\int_0^\infty \dd t \sum_x \langle \mathcal J_{0}(0) \mathcal J_x (t) \rangle_\rho 
\end{equation}
(for simplicity, we assume that the system is formally at infinite temperature, i.e.\@ the Gibbs state does not depend on the energy and is simply proportional to $\ed^{-N/\rho} = \ed^{- \sum_x |\psi_x|^2/\rho}$), 
where the current $\mathcal J_x$ is defined via $\dd |\psi_x|^2 /\dd t = \mathcal J_{x-1} - \mathcal J_x$ in $d=1$ (the definition is analogous in $d>1$ and one selects some preferential direction in eq.~\eqref{Green Kubo}).
With the current mathematical technology, one does not know how to exclude the possibility of ballistic transport (hence $\kappa (\rho) = + \infty$), even though this is very unlikely in such a system.  
However, the situation becomes much better if one assumes that, on very long time scales, the system is not entirely isolated.
As we will discuss in Section \ref{subsec: comparison DNLS}, there are at least two ways to see the DNLS system as a small perturbation of a localized system.
Here, as in \cite{Basko}, 
we first choose $J\rho/W$  (the strength of the non-linearity) to be small, and then $\epsilon := g/W$ (the hopping) to be the perturbative parameter, cfr.\@ eq.~\eqref{parameters DNLS}. 
Following a strategy successfully applied in \cite{BasileBernardinOlla}, let us assume that, on top of the Hamiltonian evolution, the dynamics has a stochastic part generated by 
$$ \mathcal L f(\psi, \overline{\psi}) \; = \; g \epsilon^n \sum_x \big(  f(\dots, -\psi_x,-\overline{\psi}_x, \dots) - f(\dots, \psi_x,\overline{\psi}_x, \dots) \big)$$
for some large power $n \gg 1$; so this dynamics randomly flips the sign of the wave vector at $x$, at a very small rate $g \epsilon^n$. 
This noise destroys the conservation of energy (but not of the norm $N$) and can be thought of as the effect of some interaction with an external medium that becomes only perceptible on very long time scales. 
What can be shown rigorously \cite{Huveneers} is that, for any $n>1$, the conductivity decays as 
$\kappa (\rho, g) \;  \le \; C_n g \epsilon^n$, as $\epsilon \to 0.$
It seems natural to think that the noise will actually favor transport, so that this result remains true for the truly isolated system (no noise). 
Though this is not precisely what is done in \cite{Huveneers}, one could also take first $g/W$ small, 
and then $\rho J/W$ as the perturbative parameter, and again show that the conductivity decays faster than any polynomial in $\rho J/W$ as this goes to 0.
A very similar result was obtained in \cite{DeRoeckHuveneers_C} for energy transport in the rotor chain.  

We would like to comment on the relevance of such results for the interpretation of some numerical results at `zero temperature'. 
In fact most of the studies related to the fate of localization in classical chains focus on the spreading of an initially localized wave packet in a medium where all other oscillators are initially at rest.
For mathematical studies, in addition to the works mentioned earlier, see \cite{BourgainWang} for the construction of quasi-periodic motion for the DNLS chain, \cite{Wayne} for the rotor chains, and 
\cite{FrohlichSpencerWayne,VittotBellissard,Poschel_1990} for other systems;  
for numerical studies, see e.g.\@ 
\cite{MulanskyPikovsky,MulanskyPikovsky_2013,LaptyevaIvanchekoFlach,IvanchenkoLaptyevaFlach} and references theirin. 
The non-linear heat equation is often used as a phenomenological model to fit the decay of the wave packet: 
\begin{equation}\label{non linear heat equation}
\partial_t \rho \; = \; \partial_x \kappa(\rho) \partial_x \rho,
\end{equation}
and it is tempting, though not obviously legitimate, to take for $\kappa$ the conductivity computed in equilibrium as in eq.~\eqref{Green Kubo}. 
If $\kappa (\rho)$ decays faster than any polynomial in $\rho$, then the spreading of the wave packet must itself  be slower than any power law in time, as can be directly deduced from eq.~\eqref{non linear heat equation}. 
It turns out however that several numerical studies report a spreading polynomial in time, e.g.\@ \cite{MulanskyPikovsky,IvanchenkoLaptyevaFlach}. 
One may think of several explanations. 
A first possibility is obviously that the true asymptotic regime is exceedingly hard to reach in such numerical experiments, and would not have been attained in numerical results predicting a power law decay. 
Another possibility is that $\kappa(\rho)$ featuring in \eqref{non linear heat equation} is not given by \eqref{Green Kubo} because the system, though spreading, does not truly have the time to reach local equilibrium; 
such a scenario does however not provide any explanation for the fast spreading. 
Finally, there is the possibility that the use of the non-linear heat equation \eqref{non linear heat equation} to model the spreading of the wave packet was simply wrong. 
Indeed, there is another way to delocalize: the initial wave packet may get dislocated in a few chunks that remain concentrated but start wandering randomly under the influence of the non-linearity. 
In this case $\rho$ does not vanish locally everywhere, and there would be no reason to think that the motion of the packet would not just be a power law in time. 
To our knowledge however, there is no explicit mention in the literature of this having been observed. 
See also \cite{Basko_2012}\cite{Basko_2014} for further discussions on this point.

\subsection{Non-perturbative origin of transport}
We follow now the approach by Basko~\cite{Basko} to estimate the conductivity in both the DNLS and the rotor system. 
The main finding of \cite{Basko} is that the dominant mechanism for transport comes from chaotic spots traveling accros the system, and modifying the energy landscape around them as they move along.   
This leads to an expression for the conductivity (for the density $\rho$) of the DNLS chain that behaves as   
\begin{equation}\label{conductivity Basko}
\kappa (\rho) \; \sim \; g \exp \Big( - C \log \frac{1}{(g/W)^p \tilde \rho} \log \frac{1}{\tilde\rho} \Big), \qquad \tilde \rho := J \rho/W, \qquad \frac{1}{2} \le p \le 3
\end{equation}
in the limit $\tilde\rho \to 0$ (the parameter $p$ could not be computed with precision), in the regime where $g \ll J \ll W$ (see Section \ref{subsec: comparison DNLS} for more details).
In \cite{Basko}, the case $d=1$ is considered exclusively. As already mentioned in the introduction, this leads to additional difficulties due to the presence of bottlenecks, 
and it is a non-trivial result of the analysis that the conductivity does not vanish. 

Here, we consider a set-up that differs from the one in \cite{Basko} in two respects. 
First we consider a dimension $d>1$, as to avoid the extra complications mentioned above.
Second, we deal with the rotor system instead of the DNLS system. 
This may slightly clarify or simplify the analysis as thermal fluctuations become then the only source of disorder.    
It turns out that in the DNLS chain, transport is eventually also primarily due thermal fluctuations, and not to fluctuations of the quenched disorder \cite{Basko}. 

We consider thus the Hamiltonian \eqref{rotors}, and we take the most favorable conditions for localization: $J \ll U \ll T$, in view of eq.~\eqref{localization rotators}.
We first will review the origin of chaos in the considered system, and describe the influence of chaotic spots on their surrounding (this is mainly standard, see e.g.\@ \cite{Chirikov}). 
We then will propose two mechanism for transport, one based on fixed chaotic spots and one on mobile spots, and we will conclude that the latter is more efficient. 
Based on this, we will obtain an expression for the thermal conductivity in the rotor system, that we finally will compare with the similar result for the DNLS chain. 

It should be clear that the theory below is based on many approximations, and that the analytical expression for the conductivity that are derived from it, as eq.~\eqref{conductivity Basko},
are exceedingly hard to validate numerically or experimentally. 
In the author's opinion, one should thus only think of all this as a proposal to understand the basic features of transport in classical systems, in the regime where localization phenomena are the most pronounced.

\paragraph{Microscopic origin of chaos.}
Let us first consider two oscillators coupled to each others: 
$$H \; = \;  \frac{U}{2} (p_1^2 + p_2^2)  - J \cos (\theta_2 - \theta_1). $$
Since the system admits two conserved quantities, the total energy $H$ and total momentum $P = p_1 + p_2$, there are only two actual degrees of freedom. 
In the variables $p = p_1 - p_2$ and $\theta = \theta_2 - \theta_1$, the Hamiltonian is that of a pendulum: 
$$H \; = \;  \frac{U}{4} P^2 + \frac{U}{4} p^2  - J \cos \theta.$$
Let us for simplicity assume $P = 0$ (this does not modify our analysis). The phase space is decomposed in four regions:
For $H = -J$, the pendulum is at rest on a stable equilibrium point; 
for $-J < H < J$, the pendulum oscillates around the stable equilibrium; 
for $H= J$, the pendulum is either on its unstable fixed point, either on the so called separatrix, i.e.\@ spends an infinite time moving towards the unstable equilibrium; 
finally for $H > J$, the velocity of the pendulum never vanishes, and it performs complete rotations.  
Unless $H=J$, the motion is always periodic, though the period diverges as the energy approaches $J$. 
This phenomenon is peculiar to classical mechanics: For a fixed number of degrees of freedom, the spectrum of generic observables becomes truly continuous on the separatrix, i.e.\@ 
$$\hat A (\nu) = \int \dd t\,  A(t) \ed^{i \nu t}$$
becomes a continuous function for initial conditions on the separatrix and for an observable $A$ that is not just a function of the energy 
(and that vanishes on the unstable fixed point so that the integral is well defined). 
Obviously, close to the separatrix, $\hat{A}(\nu)$ is still discrete but ressembles more and more a continuous function as the separatrix is approached.

The presence of `continuous spectrum' is very suggestive and indicates that already just two oscillators may behave as a perfect bath. 
It is however not straightforward to explain the emergence of chaos based only on this, and we stick here to the most conventional approach \cite{Chirikov}.  
Let us couple a third oscillator to the second one. The Hamiltonian becomes 
$$H \; = \;  \Big(  \frac{U}{4} p^2  - J \cos \theta \Big)  + \frac{U p_3^2}{2} - J' \cos \big( \theta_3 - \theta_2 \big)$$
(with actually $J' = J$, but it will turn out convenient to treat a slightly more general case). 
To analyze this system, we introduce now an approximation.   
Let us write 
$$\theta_3 - \theta_2 \; = \; \theta_3 - \frac{\Theta}{2} + \frac{\theta}{2} \qquad \text{with}\quad \Theta  = \theta_1 + \theta_2 . $$
We will assume that $\theta,p$ are the only variables that will behave chaotically, while $\Theta,P$ and $\theta_3,p_3$ still behave regularly. 
This assumption becomes surely wrong on long enough time scales, since $P$ is no longer conserved, and since the behavior of $\theta_3,p_3$ will be influenced by the chaotic behavior of $\theta,p$. 
Nevertheless, to study the shorter time scales where the chaotic behavior is generated, this approximation is reasonable and useful. 
Thus we replace the Hamiltonian $H$ by the time-dependent Hamiltonian 
\begin{equation}\label{time dependent Hamiltonian}
H(\theta,p,t) \; = \;  \Big(  \frac{U}{4} p^2  - J \cos \theta \Big)  - J' \cos (\omega t - \theta/2) \qquad \text{with}\quad \omega = \omega_3 - \frac{\omega_1 + \omega_2}{2} 
\end{equation}
with $\omega_i = Up_i$ and where we have dropped the term $U p_3^2/2$ that we treat as a constant. 

With this Hamiltonian, we can now provide a picture for chaos.
Let us assume that the instantaneous energy is very close to $J$, so that the pendulum is nearly on the separatrix. 
Most of the time, the pendulum is close to its unstable fixed point, and would move very slowly if $J' = 0$.
There the perturbation averages out over time and simply superimposes some fast oscillation. 
Sometimes though, the pendulum moves away from the unstable fixed point for a short time interval. 
When this happens, the system may absorb (or give) some energy from the time-dependent term. 
This gain or loss in energy may in turn considerably modify the time that the system will spend near the unstable fixed point before it again moves away. 
Hence, the time-dependent term of the Hamiltonian acts on the pendulum as a series of random kicks.   

Based on this picture, one may estimate how close to the separatrix the system needs to be for chaotic motion to take place, i.e.\@ to compute the width of what is called the stochastic layer. 
Let us compute the quantity of energy absorbed by the system when moving along the separatrix back and forth to the unstable fixed point. 
At $J' = 0$, the motion along the separatrix can be computed exactly: 
\begin{equation}\label{motion on separatrix}
\cos (\theta/2) \; = \;  \frac{1}{\cosh (\Omega t )} \qquad \text{with} \quad  \Omega \; = \; \sqrt{JU/2}, 
\end{equation}
where $\Omega$ is thus the frequency of the pendulum in the limit of small oscillations. 
In first approximation, the gain in energy is given by what is called the Arnold-Melnikov integral: 
$$\Delta E = \int_{-\infty}^{+\infty} \dd t \, \frac{\partial H}{ \partial t} \; = \; J' \int_{-\infty}^{+\infty} \dd t \, (\omega - \theta'(t)/2) \sin (\omega t - \theta (t)/2)   \; \sim \; J' \ed^{- |\omega|/\Omega}.$$
One may estimate the size of the stochastic layer by requiring that $\Delta E$ allows the pendulum to move back on the separatrix.
A computation yields that the surface in the $\theta,p$ plane occupied by such points is given by 
\begin{equation}\label{width stochastic layer}
w \; \sim \;  \frac{J'}{\Omega} \ed^{- |\omega|/\Omega} .  
\end{equation}

\paragraph{Resonant spots and their surrounding.}
We now come back to the Hamiltonian as given by eq.~\eqref{rotors} with more than two or three sites, actually in the thermodynamic limit. 
We assume it to be in equilibrium at temperature $T$ (and total momentum to vanish). 
We try to find what are the biggest chaotic spots in the system and we analyze their influence on their surrounding. 
While this may sound as the obvious question to ask, we will see, as we announced, that the fastest channel for transport does actually not need to involve the biggest spots. 
Hence we will keep the discussion at a minimal level.  

First we notice that closeness of the separatrix imposes at least the condition 
$|p_1 - p_2| \lesssim \sqrt{J/U}$,
i.e.\@ the oscillators 1 and 2 are close to being in resonance, cfr.~eq.~\eqref{localization rotators}. 
Second, from eq.~\eqref{width stochastic layer}, we see that the size of the stochastic layer will be maximal if $\omega \sim \Omega$.
From eq.~(\ref{time dependent Hamiltonian}-\ref{motion on separatrix}), we conclude that this will be possible if also $|p_3 - p_2| \lesssim \sqrt{J/U}$. 
Thus chaos is maximal in configurations called resonant triples in \cite{Basko}, and characterized by  
$$|p_1 - p_2| \; \sim \; |p_2 - p_3| \; \sim \; \sqrt{J/U},$$
giving rise to a width of the stochastic layer which is itself $w \sim \sqrt{J/U}$. 
The probability to observe such a spot in the Gibbs state is simply given by 
\begin{equation}\label{proba resonant triple}
\text{Proba (resonant triple)} \; \sim \; \big(\sqrt{J/U} \sqrt{U/T} \big)^2  \; = \; J/T.
\end{equation}
While both $w$ and this probability are small, there are still polynomial in the parameters of the model, which may seem contradictory with the claim that transport is non-perturbative. 
However, one must also investigate the time scales needed for energy transfer. 
We analyze two different set-ups. 

First, let us take one step back: as initially, let us just consider three oscillators, such that the two first form a separatrix and the third one introduces chaos (so no resonant triple). 
In this case, in view of the origin of chaotic motion, the time scales for energy transfer are proportional to a factor $w^{-1}$, where $w$ is the width of the stochastic layer. 
Typically, if the energy of the third oscillator is taken at random with respect to the Gibbs state, we deduce from eq.\eqref{width stochastic layer} that 
\begin{equation}\label{time scales transport}
\text{Time scale for transport} \; \sim \; Jw^{-1} \; \sim \; J \ed^{\sqrt{T/J}}.
\end{equation}
where we have dropped polynomial factors.

Next, let us consider the case where the three oscillators form a resonant triple, and let us look at the effect on a fourth one, 
but we consider the typical situation where the fourth oscillator is not in resonance with the triple. 
In this case, the time scales for transfer inside the triple are still simply polynomially small in the parameters of the model. 
Nevertheless, from Nekhoroshev estimates \cite{Nekhoroshev,Poschel_1993}, one knows that energy transfer with the outside of the spot must be exponentially small as a function of some fractional power of the small parameter of the model.  
A quick and heuristic way to understand this is to notice that the fast oscillation, with frequency $\omega_4$, of this last rotor 
as compared to the slow chaotic motion inside the spot, with frequency $\Omega$, will suppress the energy transfer exponentially as a function of $|\omega_4|/\Omega$. 
This phenomenon can be compared to the exponential suppression of the a.c.\@ conductivity as the frequency goes to infinity, in ergodic quantum systems 
\cite{arXivAbaninDeRoeckHuveneersHo_1,arXivAbaninDeRoeckHuveneersHo_2,AbaninDeRoeckHuveneers}. 
In the present case, one may model the action of the resonant triple on the 4th oscillator by a stochastic equation of the type
\begin{align*}
&\dd p_4(t) \; = \; U \dd t , \\
&\dd \theta_4 (t) \; = \; -J \sin (\theta_4 (t) - \xi(t)) \dd t, 
\end{align*}
where $\xi(t)$ is a colored noise, analytic in time, that varies on time scales of order $1/\Omega$. 
In first approximation, the rate of energy transfer is given by 
$$ \int_{-\infty}^{\infty} \dd t\,  \langle \dot\xi(t) \dot\xi(0)  \sin ( -\xi (0)) \sin(\omega_4 t - \xi(t))\rangle  \; \sim \; \Omega \ed^{- |\omega_4|/\Omega}.$$
Hence, one actually recovers an estimate of the type of eq.~\eqref{time scales transport}. 

One may iterate the above analysis and consider larger spots, with more than three oscillators in resonances. In all cases, one would find an exponential suppression of the time scales needed to transfer energy.  
The reason why we will eventually obtain that the conductivity is not exponentially suppressed, as in eq.~\eqref{conductivity Basko}, will appear clear once higher order terms will be taken into account.

\paragraph{Mechanisms for transport.}
We now move away from these microscopic considerations and introduce possible mechanisms for transport resulting from the presence of chaotic spots into the system:

a) Fixed chaotic spots produce a redistribution of the energy among the oscillators in their surrounding.
Since a finite number of oscillators may behave as a perfect bath, a spot influences in principle oscillators arbitrary far away through higher order processes. 
In an ergodic disorder free systems, it may seems odd to consider fixed chaotic spots. 
What we have here in mind is a scenario where the migration of the spots would as such not be an important effect to determine the conductivity, e.g\@ as it would occur on too long time scales. 

b) Chaotic spots travel and modify the energy landscape in their vicinity while moving around.   
The motion of chaotic spots can be understood as follows. 
To simplify the discussion, let us for the moment neglect the fact that the total momentum is conserved (this feature is non-generic and could easily be destroyed by adding some extra on-site potential). 
Let us consider three oscillators so that only the two first one are on a stochastic layer. The third oscillator may pump energy to the two first one, so as to become itself in resonance with the two first one (a resonant triple is thus created). 
At that point, one may consider rather oscillator 2 and 3 as being on the separatrix, and it is now possible for the first oscillator to pump energy to this spot and move away from the triple. 
Thus, during this process, the spot has been displaced from oscillator 1 and 2, to oscillators 2 and 3. Clearly, this can be iterated so that the spot starts exploring all the sites. 

Before proceeding further, let us insist on the fact that chaotic spots cannot just get extinct.  
This is true at both the global and local levels. 
First, we computed in eq.~\eqref{proba resonant triple} the probability of resonant triples in the Gibbs state (we could have computed the probability of simpler spots too); 
since the Gibbs state is invariant, this probability cannot evolve with time, so that the total density of spots may not vary.  
Second, spots can also not disappear or get created locally. 
Indeed, for a few oscillators, the motion outside the spots is actually known to be regular for all times, hence it is not possible that a regular motion becomes chaotic and vice versa. 

Let us now try to decide which of these two processes a) and b) above is most efficient. 
To get some sensible answer, we need to take higher order processes into account. 
Unfortunately, it is very delicate to obtain a trustable picture of the effects of higher order processes, 
and to some extent a hopeless task since there is no convergent expansion and no really obvious way to discard the terms causing the divergence 
(as an account of this difficulty, we notice that the parameter $p$ featuring in eq.~\eqref{conductivity Basko} coud not be computed explicitly in \cite{Basko}). 
We will thus simply take some cartoon that does not seem completely unreasonable. 
One may think of having performed some canonical transformation, as developed in  \cite{DeRoeckHuveneers_C,DeRoeckHuveneers_Q}, to eliminate some purely oscillatory part of the motion, leading to a Hamiltonian of the type 
\begin{equation}\label{cartoon Hamiltonian}
H \; = \;  \frac{U}{2} \sum_{x} p_x^2 - J \sum_{n\ge 1} \frac{\epsilon^n}{n!}\sum_{k_{x_1}, \dots , k_{x_n}} \cos (k_{x_1} \theta_{x_1} + \dots + k_{x_n} \theta_{x_n}),
\end{equation}
where $\epsilon$ is the perturbative parameter that, in view of eq.~\eqref{localization rotators} may be taken equal to $\epsilon = J/\sqrt{UT}$, 
where $k_{x_1},\dots k_{x_n}$ are such that $|k_{x_1}| + \dots + |k_{x_n}| = n$, and where $x_1,\dots, x_n$ form a connected set. 

We can now proceed to estimate the conductivity produced by any of the two mechanisms a) and b). 
Let us first consider fixed spots. 
Resonant triples are the most abundant and powerful chaotic spots that can show up in the system, and we may thus take them as only source of chaos in the system. 
These spots are typically separated by a distance $(J/T)^{-1} \gg 1$, according to eq.~\eqref{proba resonant triple}. 
Hence, to ensure conduction through the whole material, one needs at least to consider processes coming in the $n^{th}$ order in perturbation, with $n \sim (J/T)^{-1}$. 
Clearly this induces that transport should be suppressed as $\epsilon^{(J/T)^{-1}}$, and decays thus even faster than exponentially in the small parameter of the model. 
One should take this estimate with a grain of salt, since we may well have under-estimated the probability of resonances when taking the Hamiltonian as given by eq.~\eqref{cartoon Hamiltonian}; 
nevertheless, perturbative processes involving $n$ sites should necessarily be exponentially suppressed in $n$, hence leading to an analogous estimate.  

Let us now see that with traveling resonant spots, one obtains a rate that does actually decay subexponentially with the small parameter of the model, in a ay analogous to eq.~\eqref{conductivity Basko}. 
For this, let us simply consider two oscillators in resonance forming a separatrix 
(through higher order processes, we could actually consider separatrixes issuing from more than two oscillators but for this system this results only in a sub-dominant contribution).
Instead of coupling only one oscillator to the separatrix, we can couple many of them through higher order terms in perturbation. 
While the strength of the perturbation is made smaller, one may access smaller values for $\omega$, hence reduce the factor $\ed^{-|\omega|/\Omega}$ in eq.~\eqref{width stochastic layer}, 
i.e.\@ at the same time increase the size of the stochastic layer and reduce the time scales for Arnold diffusion (since this time scale is inversely proportional to the width of the layer).
All what needs to be done is to optimize over these two constraints: take $n$ so that the strength of the perturbation is not too small and $\omega$ is not too large. 
Given $n$ sites close to the two oscillators forming a separatrix, one may typically find $( k_{x_1},\dots , k_{x_n}) \in \{\pm 1\}^n$ so that 
$$|\omega| \; = \; |Êk_{x_1} \omega_{x_1} + \dots + k_{x_n} \omega_{x_n} |  \; \sim \; 2^{-n} U \sqrt{T/U}.$$
Hence 
$$|\omega|/\Omega \;\sim\; 2^{-n} \sqrt{T/J}  \;\ll\; 2^{-n} \epsilon^{-1} \quad \text{with} \quad \epsilon \; = \; J/\sqrt{UT}.$$
We will still under-estimate the conductivity by replacing $\sqrt{T/J}$ by $\epsilon^{-1}$, but this is irrelevant for the main message. 
Since $w$ is given by eq.~\eqref{width stochastic layer}, we look now for $n$ that maximizes
$$\frac{\epsilon^n}{n!} \ed^{-\epsilon^{-1}2^{-n}}$$
(notice that the presence of $n!$ is irrelevant for this optimization). The optimum is reached $n = \log (1/\epsilon)$, giving rise to 
\begin{equation}\label{new stochastic layer}
w \sim \ed^{- C (\log \epsilon^{-1})^2}.
\end{equation}
This width is actually much smaller than that of resonant triples but important is now that the rate for transfer is itself proportional to $w^{-1}$. 
Thus at this rate, the initial chaotic spot may generate a new spot in its vicinity, and start migrating that way in the whole system. 
We conclude that spots are distant by typically $w^{-1}$, with $w$ as given by \eqref{new stochastic layer}, and travel at a speed proportional to $w$.
Hence, as a lower bound on the thermal conductivity, we find 
\begin{equation}\label{rotor conductivity}
\kappa (T) \; \sim \; g \, \exp \Big( - C \big( \log (\sqrt{UT}/J) \big)^2\Big),  
\end{equation}
where the prefactor $g$ just stands to keep the correct dimensionality. 

Finally, we may wonder whether the whole diffusion process is not significantly slowed down by the fact that chaotic spots spend most of their time in states where they are close to the border of the stochastic layer, 
hence where the transport is significantly slower than what we have estimated. 
This effect produces indeed an extra slow-down factor, that can however be controlled by noticing that the flat measure is invariant under the flow of Hamilton's equations. 
The trajectory must hence spend the same time in all places that are accessible to it: this means thus that the frequency at which it visits places where the trajectory gets stuck must be inversely proportional to the time spent there. 
From this one sees that spots must still spend most of their time in a region where they can be considered active. The fact that some spots here and there will turn inactive for some time is basically irrelevant for the fate of the conductivity.

\paragraph{Comparison with the DNLS system.}\label{subsec: comparison DNLS}
The kind of computations above where first carried over, in much more details and at a higher level of rigor, for the DNLS systems \cite{Basko}. 
We want now to see that transport is a bit worse in the DNLS chain than in the rotor chain. 
This is visible from the fact that the logarithm features in the third power in eq.~\eqref{conductivity Basko} and only in the second power in eq.~\eqref{rotor conductivity}. 
As already alluded, there are at least two ways to consider the DNLS system as a perturbation of a localized system. 
First, which is probably the most obvious, one sees the interaction between localized modes, governed by the parameter $J$, as a perturbation. 
If one moves to the basis of the eigenmodes of the chain, the perturbation looks then as a quartic hopping of strength $J (g/W)$.
But one may also notice that, at $g=0$, the dynamics is trivially localized, and treat the hopping $g$ as the perturbative parameter. 
This choice is made in \cite{Basko}, in part because it avoids to first move to the eigenmodes basis (but this shouldn't be a fundamental choice). 
We follow \cite{Basko}, and our choice of parameters is $g \ll J \ll W$. 
Moreover, we assume again that the system is formally at infinite temperature, i.e.\@ the Gibbs state is simply proportional to $\ed^{-N/\rho} = \ed^{- \sum_x |\psi_x|^2/\rho}$.

Now one can see that the DNLS chain may be treated in essentially the same way as was the rotor chain.   
However, the presence of disorder will make the formation of separatrices much more unlikely than it was for rotors, resulting in the extra suppression factor in the expression of the conductivity. 
To see this, it suffices to consider again just two oscillators. It is convenient to move from the $(\psi_x,\bar{\psi}_x)$ coordinates to the action-angle coordinates as defined by 
$$I_x = |\psi_x|^2, \qquad \tan \theta_x = \frac{\Re \psi_x}{\Im \psi_x}.  $$ 
Again both the total energy and the total action $I_1 + I_2$ are conserved. 
Hence, writing $I = I_1 - I_2$ and $\theta = \theta_1 - \theta_2$, the Hamitlonian becomes (up to a constant)
$$H = (\omega_1 - \omega_2) I + J I^2 - g \sqrt{(I_1 + I_2)^2/4 - I^2} \cos (\theta).$$
The main difference with rotors is the presence of the linear part $(\omega_1 - \omega_2)I$; this amounts to shift the $I$ coordinate by $(\omega_1 - \omega_2)/2J$.
It may now seems that this will again lead to an exponential suppression of transport. 
Indeed, one may consider two situations. Either the disorder is typical and $|\omega_1 - \omega_2| \sim W$. 
The probability for $I$ to be of the order of $W/J$ is exponentially suppressed by the Gibbs state as $\ed^{- (W/J)/\rho}$ (we are considering the regime $\rho\ll 1$).
Or the disorder is atypical so that $|\omega_1 - \omega_2|/J \lesssim \rho$. 
However, the chaotic spots so obtained are localized on the rare places where the disorder is anomalous, 
and one knows from the above considerations that such fixed spots will also only contribute to an exponentially small conductivity.  
The way out is again to take higher order processes into account. Indeed one may create a spearatrix out of more than two oscillators. Doing so, one is able to significantly reduce the thermal cost of generating a separatrix. 
The optimization procedure used for rotors must be adapted to optimize over both the number of atoms forming the separatrix, and the number of atoms acting on a perturbation on the separatrix. 
See \cite{Basko} for a more elaborate discussion.

\section{Quantum systems}\label{sec: quantum}
We now consider quantum systems and again, we take as main examples both a quenched disordered and a translation invariant Hamiltonians, that appear as good candidates to exhibit MBL (and are known to do so in some cases). 
We explore the possibility for the two mechanisms of conduction discussed for classical chains, i.e.\@ through fixed or mobile chaotic spots, to be realized here as well. 
At first, this may not look very promising as a finite number of quantum oscillators do not host chaos. 
It is clear though that truly continuous spectrum is not needed and it is all about to know whether sparse imperfect baths will be able to play the same role as truly chaotic spots in classical systems. 
Such small baths may naturally be present as a result of disorder fluctuations or thermal fluctuations, as we will see below. 
The (imperfect) chaotic behavior will now be captured by the ETH applied to the spots, i.e.\@ we will assume that the spots are ergodic. 

The systems are as follows. First, the Hamiltonian 
\begin{equation}\label{spin chain}
H \; = \; \sum_x \omega_x \sigma_x^{(3)} \, + \,  g \sum_{x\sim y} (\sigma_x^{(1)}\sigma_y^{(1)}+ \sigma_x^{(2)}\sigma_y^{(2)} )  \, + \,  J \sum_{x,y} \sigma_x^{(3)}\sigma_y^{(3)} 
\end{equation}  
with $-W \le \omega_x \le W$ random, furnishes a prototypical example of quenched disordered system where MBL is expected to exists in $d=1$ at strong enough disorder and small enough interactions ($J \ll g \ll W$) \cite{OganesyanHuse,Imbrie_JSP,Imbrie_PRL}. 
Next, as an example of disorder free system, we consider the Bose-Hubbard Hamiltonian 
\begin{equation}\label{Bose Hubbard}
H \; = \;  U \sum_x (a_x^\dagger a_x)^2 + J \sum_{x\sim y} \big( a_x^\dagger a_y + \text{cc} \big).
\end{equation}
As for the classical rotor chain, the conditions that are most favorable for localization are $J \ll U \ll T$, assuming that the system is at equilibrium at temperature $T$. 
The possibility of realizing an MBL phase in clean systems goes back to \cite{KaganMaksimov}. 
For the particular case of the Bose-Hubbard chain (or a slightly modified version thereof), the existence of an MBL phenomenology has been proposed by 
\cite{CarleoBeccaSchiroFabrizio,DeRoeckHuveneers_Q} (see also \cite{PinoIoffeAltshuler}),
by \cite{GroverFisher,ShiulazMuller,SchiulazSilvaMueller} for systems with two species of particles, and by \cite{HickeyGenwayGarrahan} for glassy systems.

\subsection{Thermal bubbles}
For the spin chain \eqref{spin chain} in $d=1$, the most common belief is that one should distinguish three regimes, 
see e.g.\@  \cite{BAA,HuseEtAl} (see also \cite{RosMuellerScardicchio,PietracaprinaRosScardicchio} for a critical analysis of the forward approximation used in \cite{BAA}). 
For $W$ very large, the full spectrum is many-body localized; 
For intermediate values of $W$, the spectrum is separated by a many-body mobility edge: the states below some energy density remain localized while the states above become ergodic; 
Finally for $W$ small enough, all states at positive temperature become ergodic.
Another way to say this is that the critical value for $W$ separating the localized phase from the ergodic one, is expected to depend on the energy density.
This has been observed numerically for chains up to 22 atoms in several systems \cite{KjallBardarsonPollmann,LuitzLaflorencieAlet,MondragonPalHughesLaumann,SerbynPapicAbanin_2015}, 
though doubts has been raised in \cite{BeraDeTomasiWeinerEvers}. 
We notice that many-body mobility edges are properly an effect of the interaction among single particle orbitals (coming from $J\ne 0$ in \eqref{spin chain} in $d=1$) and do not exist in free systems
(since in this case the many-body energy is the sum of the energies of individual particles, and typically contains both energies below and above the single particle mobility edge, at any many-body energy density). 

For the Bose-Hubbard chain \eqref{Bose Hubbard}, one may expect a similar phenomenology with the role of low and high energies being reverted.
Indeed, close to the ground state, the transport becomes ballistic; 
At small positive temperature, there is no reason to expect any localization and the system is most likely to be ergodic;  
Finally, at high temperature, there is typically an important energy mismatch between near sites, and a many-body localized phase could emerge.
Numerical evidences for the existence of an MBL phase in disorder free systems are not conclusive \cite{SchiulazSilvaMueller,HickeyGenwayGarrahan}. 
Moreover, in \cite{PapicStoudenmireAbanin}, the authors insist on the importance of finite size effects in detecting MBL in clean systems. 
They identify that, in the thermodynamic limit, the spectral lines of the unperturbed systems must get broaden and overlap each other. This furnishes a practical criterion to analyze whether or not the thermodynamical limit is reached. 
Based on this, they actually conclude that there is no MBL phase in the clean system that they have investigated. 

We now review the argument in \cite{DeRoeckHuveneers_Scenario,DeRoeckHuveneersMuellerSchiulaz} showing that a transition between an ergodic and an MBL phase as a function of a thermodynamical parameter is impossible. 
This rules out both the existence of many-body mobility edges in disordered systems, and the existence of an MBL phase in clean systems (since we are not aware of any clean system which full spectrum could be localized). 
While these two systems are here treated on equal foot, it remains unclear from this point of view why the upshot of numerical experiments are so different in disordered and clean systems. 

For the sake of concreteness, let us consider the spin chain \eqref{spin chain} and let us assume that the disorder is intermediate so that high energy states are ergodic, while low energy states exhibit at least some MBL phenomenology. 
The basic idea is that, for typical states at any temperature, there are are thermal fluctuations implying that, locally, the state may be at a different temperature than the global temperature
(in the same way as the air in a room is not precisely at a constant density at the mesoscopic level).  
Hence, at any energy density, there is always a fraction of the space occupied by spots where the state is actually thermal inside the spot
(the situation is actually slightly more complicated since disorder fluctuations may actually forbid some spatial regions to become thermal but,
while this fact needs definitely to be taken into account, especially in $d=1$, we will ignore it here to focus on the core of the argument; eventually it turns out that this modifies transport properties but does not stop transport). 
Such hot thermal bubbles behave as baths and will play the role of chaotic spots in classical systems: they travel accros the system and provide a mechanism for transport.

In the above description, we did not argue that the imperfect bath made out of the thermal bubble is good enough to effectively play the role of a chaotic spot.  
This requires a more quantitative argument.  
While we have until now looked at the system dynamically, it is more convenient to formulate this argument for the eigenstates of the system. 
The eigenstates of a localized system should have an approximate product structure, as can e.g.\@ be seen from the fact that their entanglement entropy follows an area law. 
Here we will show that the presence of thermal bubbles forces product states to strongly hybridize, so that this structure should entirely be lost for the eigenstates.

Let us work in $d=1$ for convenience, and let us divide the system into boxes of a few sites each, say $n$. 
The main interest of this coarse grained description is that the coupling between each box is small as compared to the energy in the bulk of the boxes, so that it makes sense to say that a box is at low or high energy. 
Let us then remove all the couplings between the boxes, and let us consider an initial product state that is an eigenstate of this uncoupled system.
So each box is either in a thermal state (if the energy is high) or a localized state (if the energy is low); the presence of some possible intermediate states is irrelevant. 
At very small temperature, it is very well possible that the huge majority of the boxes is in a localized state, but if the initial state was chosen typical, there is always also a small fraction of boxes in a thermal state.     
To clearly see hybridization, let us introduce an additional large number $N$ and let us declare that a region is thermal if $N$ boxes near each other are in a thermal state. 
Again, all typical states have a fraction of thermal regions.
Within a thermal region, we then re-insert the coupling among boxes, and we consider an ergodic state inside this region. 
 
We now show that, upon re-inserting the coupling among boxes not inside a thermal region, 
any state $|\eta\rangle$ with a thermal region occupying boxes $a, a+1, \dots , a+N-1$ must hybridize with some state $|\eta' \rangle$ with a thermal region on boxes $a+1, a+2, \dots , a+N$. 
Let us show that $|\eta\rangle$  hybridizes with some state $| \eta'' \rangle$ with a slightly larger thermal region concentrated on  the boxes $a,a+A, \dots , a+N$; 
the hybridization of $|\eta'' \rangle$ with $|\eta'\rangle$ is then seen in the same way.  
We need to compare the matrix element of the perturbation with the level spacing. 
We may choose $|\eta''\rangle$ so that the level spacing is $\delta \sim 2^{-nN}$ (for simplicity we work here at infinite temperature, in general $2$ should be replaced by $\ed^{s(T)}$ with $s(T)$ the entropy density at temperature $T$).
Denoting by $V$ the perturbation among blocks, we have from ETH that $\langle \eta Ê|ÊV |Ê\eta'' \rangle \sim 2^{-nN/2}$, hence 
$$\frac{|\langle \eta |ÊV |Ê\eta'' \rangle | }{\delta} \; \sim \; 2^{nN/2} \; \gg \; 1,$$
which is a sufficient condition for hybridization.  

At this point, it is clear that this hybridization procedure can be iterated so that the product structure is completely destroyed. 
We refer to \cite{DeRoeckHuveneersMuellerSchiulaz} for the study of several caveats in the above argument.  
Finally, we notice that the conductivity due to this mechanism vanishes drastically in the low temperature limit. 
Indeed, the probability of finding a bubble should decay as $\ed^{- \varepsilon (1/T - 1/T_c)}$ as $T\to 0$, where $T_c$ is the `critical' temperature above which the material becomes ergodic in an obvious way 
(i.e.\@ the temperature of the putative mobility edge), and where $\varepsilon$ is the typical amount of energy in a bubble. 
Hence the thermal conductivity $\kappa (T)$ should itself become proportional to $\ed^{- \varepsilon /T }$ in the limit $T\to 0$.  
As in the classical case, one may argue that this neglects the fact that a bubble may spend a huge time in states that are nearly frozen (by dissipating instead of travelling). 
However, the mechanism of conduction does not require bubbles to become larger as the temperature is decreased, hence this should at most result in a temperature independent pre-factor.  
This reasoning is actually valid only in $d>2$. 
Indeed, in $d=1,2$, isolated bubbles become weakly localized, and interactions among bubbles become necessary for diffusion to take place. 
Thus in $d=1,2$ the above computation over-estimates the value of $\kappa (T)$ in the $T\to 0$ limit. 
See also \cite{GornyiMirlinMuellerPolyakov}.

\subsection{Quenched ergodic spots}

Finally we consider strongly disordered quantum systems where there are no thermal bubbles and we ask whether quenched ergodic spots, resulting from disorder fluctuations, may cause transport.
At any value of the disorder, there is a density of such spots, though their proportion goes to zero as the disorder increases. 

Let us assume that, between the spots, the material is perfectly localized and is described by local integrals of motion (LIOM) with a localization length $\xi \ll 1$, while the dynamics inside the spots obeys ETH. 
We imagine having turned off the coupling between the spots and the localized medium, and want to understand the influence of the spots on the MBL material when restoring the coupling. 
One may adopt two extreme attitudes. 
First, to decide whether or not a LIOM needs to be thermalized by the spot, one may compare the effective coupling, that decays as $\ed^{-\ell/\xi}$ where $\ell$ is the distance of the LIOM to the spot, 
to the level spacing inside the spot, which is roughly $\delta = 2^{-L^d}$ (at maximal entropy), where $L$ is the linear size of the spot and $d$ the spatial dimension. 
Based on this view, one would conclude that  only LIOMs that are at a distance smaller that $\ell_c \lesssim \xi L^d \log 2$ are thermalized, hence that localization survives as long as the typical distance between spots is larger than this length. 
However, one quickly realizes that this criterium is not very consistent, since the LIOMs that get thermalized become themselves part of the ergodic spot, hence reduce the value of the level spacing.
The other extreme attitude is then to take for the level spacing $\delta = 2^{- (L + \ell_c)^d}$, corresponding to the level spacing of the system made of the original spot and the spins that get thermalized. 
One obtains thus a self-consistent equation for $\ell_c$: 
\begin{equation}\label{lc d1}
\ell_c \; = \; \frac{\xi \log 2}{2} (L + \ell_c)^d
\end{equation}
(the factor $1/2$ comes from the fact that we assumed that matrix elements scale as $\ed^{-\ell_c/\xi} 2^{-(L+\ell_c)^d /2}$, consistently with the use of random matrix Ansatz).
In $d=1$, replacing $L+\ell_c$ by $L+2\ell_c$ in eq.~\eqref{lc d1} to take account of the fact that the spot is surrounded by spins on both sides, we get
\begin{equation}\label{lc d2}
\ell_c \; = \; \frac{1}{2} \frac{\log 2}{\xi^{-1} - \log 2} \, L,
\end{equation}
which remains finite provided that $\xi < (\log 2)^{-1}$. 
In dimension $d>1$, there is no solution as soon as $L\gtrsim \xi^{-1/(d-1)}$ as $\xi \to 0$ and, since there are arbitrarily big spots in the thermodynamic limit, we would conclude that MBL is not stable in $d>1$. 
The same argument applies obviously to systems with sub-exponentially decaying interactions in all dimensions \cite{Burin,Yao_EtAl}. 
We notice that a similar distinction (weak and strong localization) was already observed in \cite{NandkishoreGopalakrishnanHuse}, for an MBL system weakly coupled to a bath.

In \cite{DeRoeckHuveneers_2016}, the authors propose a description of the interface region near the core of the ergodic spots. 
One of the main consequence of their analysis is that the expressions (\ref{lc d1}-\ref{lc d2}) hold, though they are no longer based on the crude assumptions made above.
In \cite{DeRoeckHuveneers_2016}, random matrix, or ETH, is used as much as possible, taking into account energy conservation as the only constraint.  
ETH furnishes an Ansatz for the elements of a local operator $A$: 
\begin{equation}\label{ETH operators}
\langle E' |ÊA  |ÊE\rangle \; = \; \sqrt{\delta} \, \sqrt{v(\omega)} \, \eta_{E,E'}, \qquad \omega = E' - E  
\end{equation}
where $\delta$ is the full level spacing, $v(\omega) = \int_0^\infty \dd t \, \langle A(t) A(0) \rangle_E \ed^{-i \omega t}$ is the dynamical structure factor (and $\int \dd \omega \, v(\omega) \sim 1$), and $\eta_{E,E'}$ is a random phase. 
For systems with local interactions, $v(\omega)$ is typically constant for $\omega$ of the order of the energy per site, then decays exponentially.
In the present case instead, the structure factors will acquire more and more `structure' as long as the distance to the bath increase; 
hence they will be the key to describe the boundary region. 

Let us assume a bath of linear size $L$ to which are coupled spins up to some distance $\overline{\ell}$. 
Let us assume that this full region is described by ETH, i.e.\@ local operators everywhere are described by eq.~\eqref{ETH operators} 
(as we will see, if $\overline{\ell} > \ell_c$ with $\ell_c$ as predicted by eq.~(\ref{lc d1}-\ref{lc d2}), this hypothesis turns out to make no sense, hence we actually assume $\overline{\ell} \le \ell_c$). 
Obviously the level spacing $\delta$ is the same for all of them and corresponds to the level spacing of the full system. 
Instead, as a consequence of energy conservation, the analysis of \cite{DeRoeckHuveneers_2016} reveals that  the structure factors become more and more concentrated on a few spikes as one goes away from the bath. 
More precisely, the structure factor of an operator at distance $\ell$ is essentially non-zero only for $\omega$ in some windows which total length is of the order of  $(g/W)^\ell W = \ed^{- \ell/\xi} W$, 
where $W$ is the disorder strength (typical energy per site).
Obviously, the Ansatz \eqref{ETH operators} looses its meaning once this size becomes comparable to the level spacing. 
This furnishes a criterium to know when spins stop getting hybridized, and one recovers eq.~(\ref{lc d1}-\ref{lc d2}).

Let us mention that the `size of the support of the structure factor' is not a vague notion. It can be measured through IPR of a local operator $O_x$, as defined by 
$$\mathcal D (O_x,E) \; = \;  - \log  \sum_{E'} |\langle E' |ÊA |ÊE \rangle |^4.$$
for some given energy $E$, that we here assume to be in the middle of the band (so we drop $E$ from our notations).
For an ergodic system, $\mathcal D(O_x)\sim V \log 2$, and for an MBL system $\mathcal D (O_x) \sim 1$. 
Let us now consider the case of an ergodic spot coupled to an MBL region, and let us denote by $V_{th}$ the number of sites in the original spot or at a distance smaller than $\ell_c$ from the spot, with $\ell_c$ as in eq.~(\ref{lc d1}-\ref{lc d2})
(so possibly $V_{th}$ is the full volume). The behavior of $\mathcal D(O_x)$ predicted by \cite{DeRoeckHuveneers_2016} is 
\begin{equation}\label{ipr predictions}
\mathcal D(O_x) \; \sim \; V_{th} \log 2 - \frac{2}{\xi}  \ell, 
\end{equation}
assuming that $x$ is at distance $\ell \le \ell_c$ from the spot ($\ell=0$ for $x$ in the spot). 

It is worth spelling out the two aspects contained in eq.~\eqref{ipr predictions}: 
First, all the spins in the cross region maximally participate to the effective dimension of the bath for an operator inside the ergodic spot, since $\mathcal D(O_x)$ is then proportional to the full $V_{th}$. 
Second, $\mathcal D(O_x)$ decays linearly with the distance from the bath for operators inside the cross-over region; hence, if the size of the spot is sub-critical (i.e.\@ if $\ell_c < + \infty$), 
the parameter $\mathcal D(O_x)$ describes a smooth transition between the core of the spot and the MBL region.  

We refer again to \cite{DeRoeckHuveneers_2016} for the discussion of possible caveats in the above analysis, 
in particular related to the back reaction of the localized system on the bath and the discussion of various `proximity effects' \cite{NandkishoreGopalakrishnan,Nandkishore,HuseNandkishorePietracaprina}. 
Finally, let us say a word about time scales for transport. Let us assume $d>1$ so that the system is diffusive. 
However, the thermal conductivity vanishes drastically as the bare localization length $\xi\to 0$, i.e.\@ the localization length in absence of super-critical spots. 
Indeed the distance between spots decays as $dist \sim \ed^{1/\xi}$ and the time for thermalization as $t_{th} \sim g^{-1} \ed^{dist/\xi}$, where $g$ is the bare hopping constant. 
This yields $\kappa \sim g \exp(- \ed^{1/\xi}/\xi)$. 

\subsection*{Acknowledgements}
I thank Denis Basko for a critical proofreading of this manuscript. 
I thank Wojciech De Roeck, Markus M\"uller and Mauro Schiulaz for collaboration and discussions on most of the material covered in this review. 
I thank the CNRS D\'efi InPhyNiTi grant `MaBoLo' for financial support.

\bibliographystyle{plain}
\bibliography{MBL_Review.bib}

\end{document}